# Widely tunable 2 µm hybrid laser using GaSb semiconductor optical amplifiers and Si$_3$N$_4$ photonics integrated reflector

**NOUMAN ZIA\*, SAMU-PEKKA OJANEN, JUKKA VIHERIALA, EERO KOIVUSALO, JOONAS HILSKA, AND MIRCEA GUINA**

*Optoelectronics Research Centre, Physics Unit, Tampere University, Korkeakoulunkatu 3, 33720*
*\*Corresponding author: nouman.zia@tuni.fi*

Tunable lasers emitting at a 2 – 3 µm wavelength range and compatible with photonic integration platforms are of great interest for sensing applications. To this end, combining GaSb-based semiconductor gain chips with Si$_3$N$_4$ photonic integrated circuits offers an attractive platform. Herein, we exploit the low-loss features of Si$_3$N$_4$ waveguides and demonstrate a hybrid laser comprising a GaSb gain chip with an integrated tunable Si$_3$N$_4$ Vernier mirror. At room temperature, the laser exhibited a maximum output power of 15 mW and a tuning range of 80 nm (1937 – 2017 nm). The low-loss performance of several fundamental Si$_3$N$_4$ building blocks for photonic integrated circuits is also validated. More specifically, the single-mode waveguide exhibit transmission loss as low as 0.15 dB/cm, the 90° bend has 0.008 dB loss, and the 50/50 Y-branch has an insertion loss of 0.075 dB.

1. Introduction

Photonics integration has been widely recognized as a major driving force fostering the development of new photonics applications, including data center transceivers, wearable sensors for health monitoring, or solid-state LIDAR solutions. In fact, tailored multi-functional photonics integrated circuits (PICs) are the key building blocks enabling many new applications. PICs bring important benefits for volume scaling at affordable costs, miniaturization, and improved reliability. These benefits have been widely exploited by combining silicon photonics with InP-based gain chips for complex high-bandwidth optical transceivers [1,2] operating at telecom wavelengths. While this established PIC platform can find use in other applications, it can provide marginal coverage for the 2 – 3 µm wavelengths window required for sensing applications, such as atmospheric pollutants [3,4] or real-time monitoring of glucose levels [5], for example. These types of applications require lasers able to investigate multiple complex spectral fingerprints spanning over a broad spectral band [6].

To date, PIC-based tunable hybrid external cavity lasers near the 2 µm range are largely based on a submicron silicon-on-insulator (SOI) platform [7–9]. This limits the extension of the technology towards 3 µm due to high mode leakage to lossy SiO$_2$ bottom and top claddings [10]. Moreover, the limited performance of the InP material system in this wavelength range has restricted the development of InP/SOI PICs up to 2.35 µm [9]. Alternatively, silicon nitride (Si$_3$N$_4$) has been hailed as a promising integrated photonic platform [11] offering ultra-low propagation losses, negligible nonlinear absorption, and a wide transparency window extending from visible to mid-IR. This platform offers low index contrast between Si$_3$N$_4$ waveguide material and SiO$_2$ cladding, which is a benefit allowing higher fabrication tolerances for waveguide circuitry. Furthermore, on-chip sensors relying on an evanescent coupling scheme become more sensitive when the low contrast between waveguide and air enables higher interaction between propagating light and environment or functionalized surfaces. Then in terms of III-V gain materials matching the spectral needs, the GaInAlAsSb/GaSb-based type-I laser diodes have excelled in the 2 – 3 µm wavelength range [12–14] ensuring high gain at relatively low thresholds and low operation voltage, in particular around 2 µm.

In this paper, we report for the first-time the use of a $Si_3N_4$-PIC Vernier reflector for locking and tuning the wavelength of a 2 µm hybrid laser incorporating a GaSb type-I quantum wells heterostructure. We deployed a relatively thick (800 nm) $Si_3N_4$ platform preventing the mode overlap with the claddings. The demonstrated hybrid laser shows the potential of this platform for chip-scale tunable lasers at 2 µm wavelengths and beyond. In particular, a record wide tuning range of 80 nm (1937 nm – 2017 nm) and a relatively high output power of 15 mW are demonstrated for room temperature operation. We further discuss the design of $Si_3N_4$ Vernier PIC and explore in detail the loss performance enabling further wavelength scaling.

## 2. Design and hybrid integration

Figure 1 shows a detailed schematic diagram of the integrated hybrid laser, which consists of a 1mm long ridge waveguide (RWG) reflective semiconductor optical amplifier (RSOA) end-fire coupled with a $Si_3N_4$ Vernier PIC. The RSOA heterostructure was grown via molecular beam epitaxy (MBE) on a (100) n-GaSb substrate. The active region consists of 10 nm thick In(0.25)GaSb type-I double quantum wells embedded in lattice matched 260 nm Al(0.25)GaAsSb waveguide and 10 nm barrier layers. The waveguide is sandwiched between 2000 nm thick p- and 2700 nm thick n-Al(0.50) GaAsSb claddings. The heterostructure was processed into 5 µm wide ($W_{RSOA}$) and 2076 nm deep ($t_{RSOA}$) RWG geometries targeted for single transverse mode operation. The coupling facet of the RSOA RWG was tilted 7° and anti-reflection (AR) coated to suppress lasing in the RSOA and to obtain a broad spectrum.

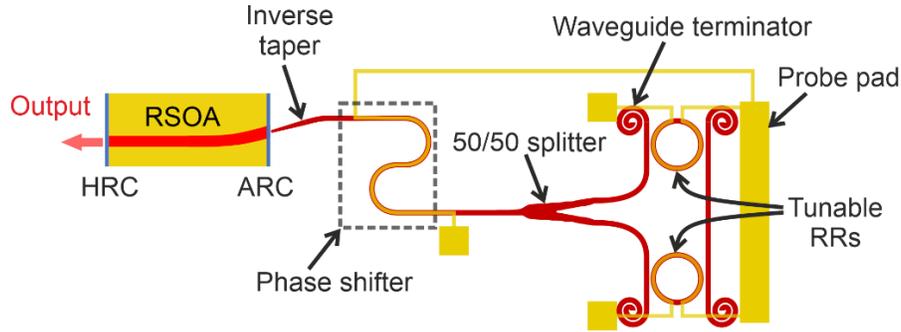

**Fig. 1.** Schematic of the widely tunable laser design showing the RSOA gain chip and $Si_3N_4$ Vernier mirror together with important PIC building blocks.

The Vernier PIC was realized on LIGENTEC 800 nm thick $Si_3N_4$ platform through open-access foundry services [15] and each component was optimized using Ansys Lumerical software suite [16] for a low-loss, broadband operation around 2 µm. The 1 µm wide strip waveguides were designed for low-loss (0.05 dB/cm) single transverse mode operation. The PIC coupling interface consists of an AR-coated waveguide with an inverse taper edge coupler to maximize the coupling efficiency. The inverse taper is tilted at 17° to match the output angle of RSOA. An ultra-low loss (~0.05 dB), broadband (> 200 nm), and small footprint (34 µm x 7 µm) 50/50 Y-branch splitter is designed through Ansys Lumerical Photonic Inverse Design (PID) package to split the light into two arms. Each arm includes a ring resonator (RR), forming a single-pass loop mirror. The light travels through each RR with slightly different radii, which leads to the Vernier effect [17]. In this design, the RR radii were chosen to be 100 µm and 96.7 µm, which corresponds to the 3.03 nm and 3.13 nm free spectral ranges (FSRs) [18] for a group index of 2.09. Figure 2 (a) shows the simulated transmission response for two RRs with a coupling coefficient of 0.04 between the ring and bus waveguide and no waveguide loss. The simulated Vernier transmission, in Figure 2 (b), which is the overlap between the spectral response of both RRs gives rise to an FSR of 90 nm for our design. Metal heaters were added on the ring waveguides to tune the Vernier transmission covering the full FSR. A thermal phase shifter (PS) was also added in the PIC to align cavity resonance with the Vernier resonance, i.e., the $Si_3N_4$ PIC and RSOA are perfectly phase-matched. Spiral waveguides were placed at the through ports of each RR to suppress the back reflections. The waveguide bends on the PIC were designed for a negligible loss of 0.003 dB/90°, which corresponds to a radius of 100 µm.

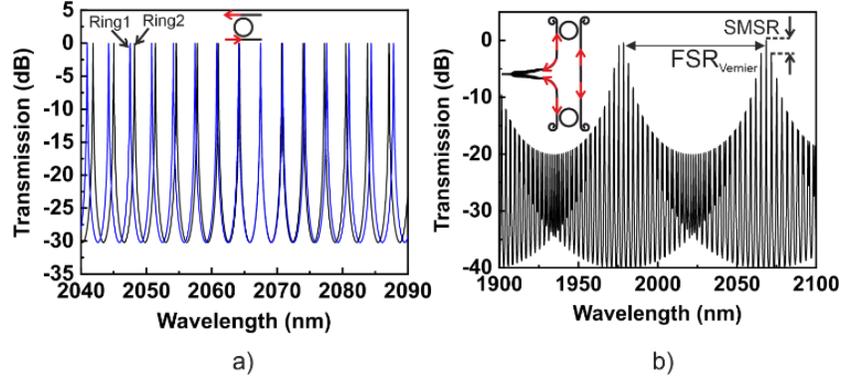

**Fig. 2.** Theoretical transmission spectrum of (a) two RRs with slightly different FSR, and (b) the Vernier filter.

The RSOA-PIC coupling interface consists of an inverse taper edge coupler to maximize the coupling efficiency. A schematic of the coupling interface between RSOA and a $Si_3N_4$ PIC with an inverse taper is shown in Figure 3. The cross-sectional optical field simulations performed with Ansys Lumerical Finite Difference Eigenmode (FDE) are also shown in Figure 3. Simulation represents the fundamental transverse electric (TE) modes of the RSOA waveguide, the $Si_3N_4$ waveguide at the taper input, and the $Si_3N_4$ waveguide at the taper output, with their mode field diameters (MFDs) and effective index ($n_{eff}$).

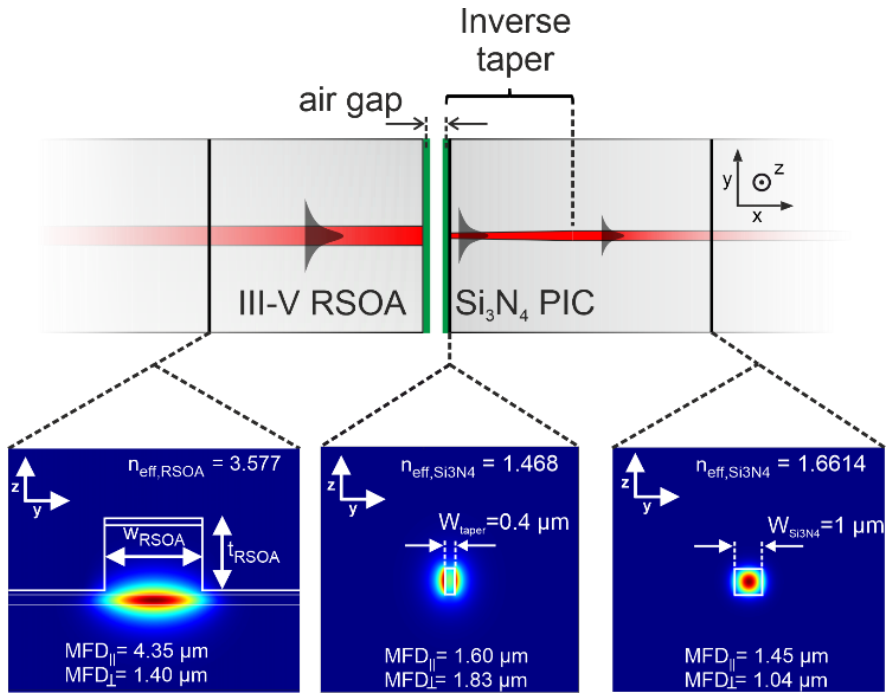

**Fig. 3.** Schematic of the edge-coupling interface between RSOA gain chip and $Si_3N_4$ PIC, together with simulated 2D fundamental TE mode profiles of RSOA waveguide ($W_{RSOA}$ = 5 µm, $t_{RSOA}$ = 2076 nm), the $Si_3N_4$ waveguide at the taper input ($W_{taper}$ = 0.4 µm), and the $Si_3N_4$ waveguide at the output ($W_{Si3N4}$ = 1 µm). $MFD_\perp$ = mode field diameter vertical, $MFD_\parallel$ = mode field diameter horizontal.

To optimize the taper input width, simulations were performed with the Ansys Lumerical FDE solver. Figure 4 (a) shows the simulated mode mismatch loss between fundamental TE modes of RSOA and $Si_3N_4$ taper input for fixed RSOA geometry and varying the $Si_3N_4$ taper input widths. A minimum mode mismatch loss of ~0.85 dB is obtained for 0.4 µm wide taper input width. The optimum taper length was chosen by simulating the coupling loss between taper input and output waveguide modes as a function of taper length. These simulations were performed using the Ansys Lumerical Eigenmode expansion (EME) solver and results are shown in Figure 4 (b) for three different input taper widths. The results show that a taper length of 100 µm is sufficient for < 0.1 dB coupling loss when the input taper width is 0.4 µm. The propagation loss is much higher for narrow taper widths since the optical mode overlaps strongly with the SiO$_2$ claddings over propagation through taper length. Therefore, the total coupling loss for edge couplers with narrow input taper widths is higher.

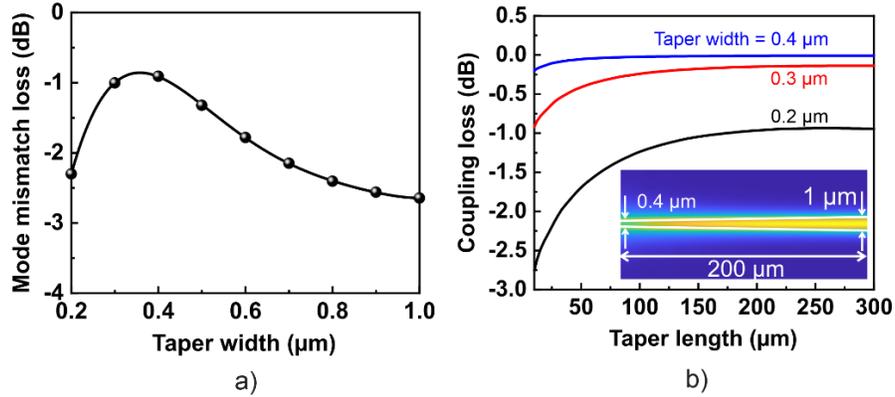

**Fig. 4.** (a) Simulated mode mismatch loss between RSOA and Si$_3$N$_4$ taper input waveguide as a function of taper input width, and (b) simulated coupling loss between input and output taper waveguide as a function of taper length. The inset in (b) shows the simulated optical field in 200 µm long taper for 0.4 µm input taper width.

To ensure stable single-mode operation the power coupling coefficient was chosen carefully by simulating the Vernier response as a function of the RR coupling gap. The component level simulation for ring directional coupler, bends, and waveguides were performed using the Ansys Lumerical FDTD and FDE solvers, and their response was imported to Interconnect for circuit level simulation of the Vernier filter. The simulation results are shown in Figure 5. The single mode suppression ratio (SMSR) (absolute value) increases, whereas the linewidth decreases with increasing coupling gaps. By increasing the coupling gap the effective cavity length of RR is extended which makes the photon lifetime much longer [19]. Therefore, the linewidth of each RR is reduced resulting in an overall improved Vernier linewidth and SMSR. The improvements come with an increase in transmission loss due to propagation loss caused by a large number of round trips in RR. Therefore, the choice of coupling gap is based on the balance between spectral purity and output power of the laser. In this work, coupling gaps of 700 nm and 800 nm were used for two Vernier PICs, while the measurements were performed only for the 800 nm Vernier PIC.

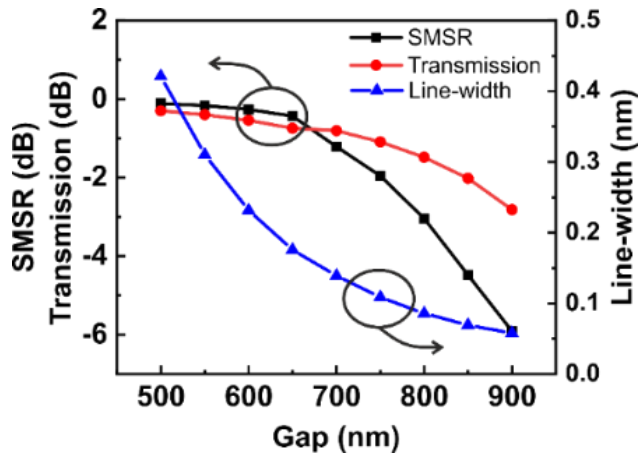

**Fig. 6.** Simulated Vernier filter SMSR, transmission, and linewidth as a function of the coupling gap between ring resonator and bus waveguide.

### 3. Experimental results

The propagation loss of the strip waveguide was measured by the cut-back method, where a set of three spirals with different lengths and uniform bend radii of 100 µm were fabricated. The loss of these spirals was measured with respect to a straight waveguide reference using a Norcada DFB laser at 2 µm wavelength. The waveguide loss as a function of spiral length is shown in Figure 6 (a), where the inset shows a simulated single TE mode of strip waveguide. Fitting a linear curve yields a waveguide propagation loss of 0.15 dB/cm, which is about 30 % larger than the simulated propagation loss (0.05 dB/cm) without scattering effects. The y-intercept of the linear fit gives a 0.008 dB/90° bend loss, which is 37 % larger than the simulation estimate. To characterize the Y-branch splitter, cascades of 15 and 30 Y-branches in series were fabricated and their loss was measured with respect to a straight waveguide. By fitting a linear function to the measured loss values an insertion loss of 0.075 dB was measured at 2 µm, which is close to the simulated (0.05 dB) value. The inset in Figure 6 (b) shows the simulated field profile of an optimized Y-branch designed through PID and fabricated in this work.

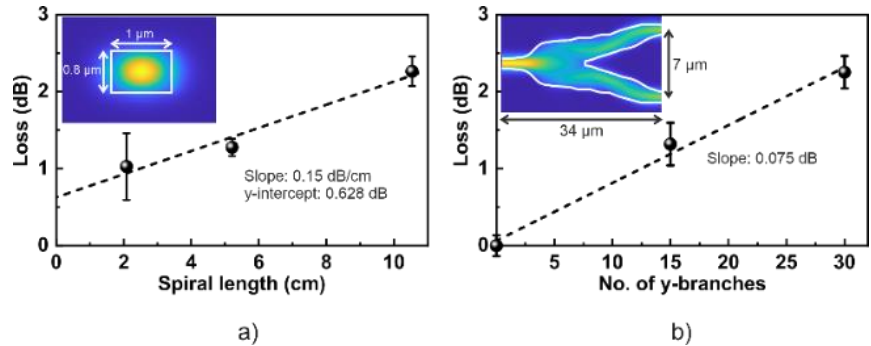

**Fig. 6.** (a) The measured loss of spiral waveguides of different lengths with respect to a straight waveguide, and (b) the measured insertion loss of the number of Y-branches. The inset in (a) shows a simulated single transverse mode field in strip waveguide, and in (b) shows the simulated field profile in PID optimized Y-branch.

To realize the GaSb/$Si_3N_4$ hybrid laser, a 1 mm long RSOA gain chip was p-side down mounted on an aluminum nitride (AlN) submount which was stabilized at 23 °C using a thermoelectric temperature control system. RSOA and Vernier PIC were brought as close as possible for efficient end-fire coupling. The continuous wave (CW) current was injected into the RSOA with probe needles, and the output of the hybrid laser was coupled to a MM fiber, which was connected to a photodiode. The measured light-current (LI) curve is shown in Figure 7 (a). The laser emits up to 15 mW CW output power and has a threshold current below 100 mA. The strong kinks in the LI curve are explained by the phase-matching oscillation between RSOA and $Si_3N_4$ cavities formed when the phase of the RSOA changes with the input current. By tuning the intracavity thermal PS for each input current an efficient phase matching can be achieved. The output power and threshold current is expected to be improved by changing the taper input width (0.2 µm) used in this work to 0.4 µm. The spectra measured at different gain currents in Figure 7 (b) show the wavelength locking at the current as low as 100 mA with SMSR up to 25 dB.

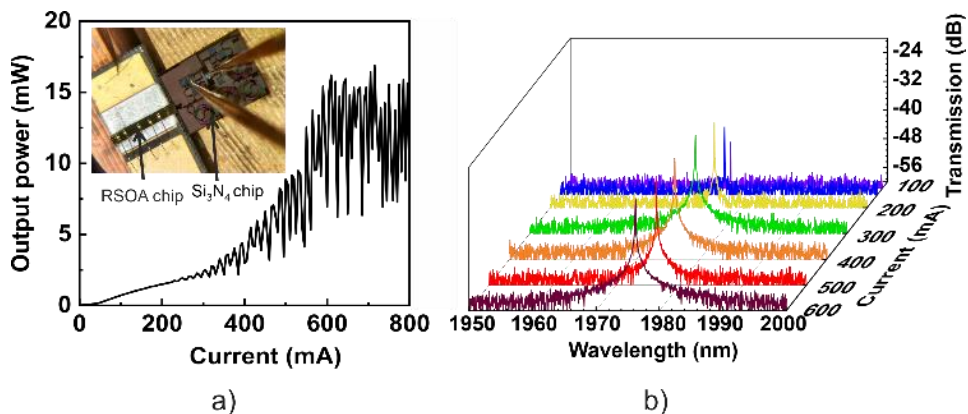

**Fig. 7.** Measured a) output power, and b) spectra of hybrid laser at different CW injection currents. Inset in (a) gives a top camera view of the edge-coupled GaSb RSOA gain chip and $Si_3N_4$ PIC.

The emission wavelength is tuned by driving the thermal heater over one of the RRs. Figure 8(a) shows the superimposed laser emission spectra achieved by tuning only one RR up to 350 mW drive power. The current injected in the RSOA was kept at 600 mA during the measurements. Changing the temperature of one RR shifts its transmission spectrum and thus the overlapping wavelength jumps from one transmission peak of the unheated RR to the next. The spacing between these peaks is found to be ~3.11 nm, which corresponds to the FSR of the unheated ring. The laser shows a tuning range of over 80 nm covering the wavelengths between 1937 nm to 2017 nm, which is the highest reported for hybrid Vernier lasers around 2 µm wavelength. The dependence of the lasing wavelength on the power dissipated in the heater is shown in Figure 8(b). Wavelength scanning across the tuning range of 80 nm is obtained with a heater power consumption of ~ 363 mW. The SMSR corresponding to each lasing wavelength is also shown in Figure 8(b). It can be seen that the laser exhibits an SMSR of more than 20 dB SMSR across the entire tuning range.

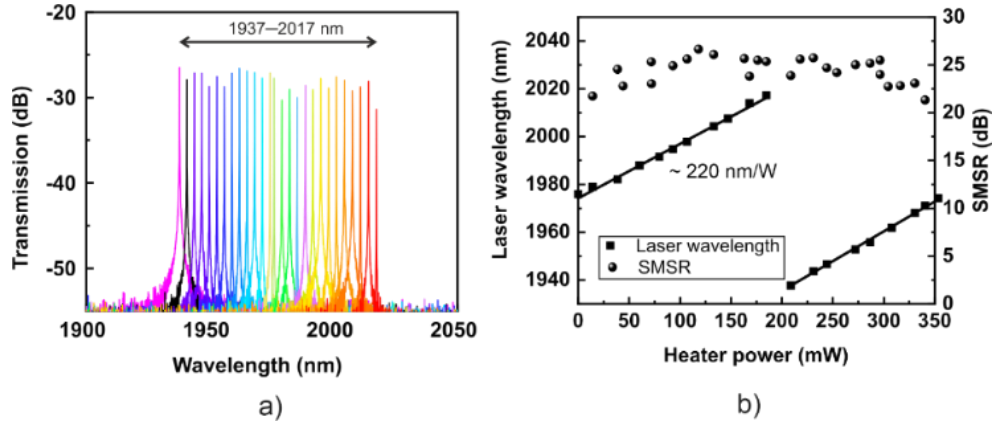

**Fig. 8.** a) Superimposed emission spectra of laser tuned by varying the RR heater power, and b) wavelength tuning as a function of heater power, and SMSR in the entire tuning range.

## 4. Conclusion

In conclusion, we have demonstrated for the first time a widely tunable GaSb/$Si_3N_4$ hybrid laser around the 2 µm wavelength. The laser employs the Vernier mechanism between two thermally tunable RRs for wavelength filtering and tuning. The hybrid laser exhibited a relatively high CW output power of 15 mW, a low threshold current of 100 mA, and a broad tuning range of 80 nm covering the wavelengths between 1937 nm to 2017 nm. The Vernier PIC is based on ultra-low loss $Si_3N_4$ building blocks measured at 2 µm. In fact, we measured 0.15 dB/cm propagation loss for the straight waveguides, 0.008 dB/90° bend loss, and 0.075 dB for Y-branch insertion loss. In the future, the performance of the hybrid laser will be improved by increasing the coupling efficiency through an optimal taper geometry. Finally, the emission wavelength will be extended beyond 2 µm to cover the needs of PIC-based hybrid lasers for sensing applications at the 2 – 3 µm range.

**Funding.** The research was funded by EU Business Finland projects RAPSI (decision 1613/31/2018) and PICAP (decision 44761/31/2020). The work is also part of the Academy of Finland Flagship Programme PREIN #320168.

**Acknowledgments.** The authors wish to thank MSc. Jarno Reuna for preparation of AR/HR coatings and Ms. Mariia Bister for wafer-level fabrication.

**Disclosures.** The authors declare no conflicts of interest.

**Data availability**. Data underlying the results presented in this paper are not publicly available at this time but may be obtained from the authors upon reasonable request.

**References**

1. R. Jones, P. Doussiere, J. B. Driscoll, W. Lin, H. Yu, Y. Akulova, T. Komljenovic, and J. E. Bowers, IEEE Nanotechnol. Mag. **13**, 17 (2019).
2. S. Fathololoumi, D. Hui, S. Jadhav, J. Chen, K. Nguyen, M. N. Sakib, Z. Li, H. Mahalingam, S. Amiralizadeh, N. N. Tang, H. Potluri, M. Montazeri, H. Frish, R. A. Defrees, C. Seibert, A. Krichevsky, J. K. Doylend, J. Heck, R. Venables, A. Dahal, A. Awujoola, A. Vardapetyan, G. Kaur, M. Cen, V. Kulkarni, S. S. Islam, R. L. Spreitzer, S. Garag, A. C. Alduino, R. K. Chiou, L. Kamyab, S. Gupta, B. Xie, R. S. Appleton, S. Hollingsworth, S. McCargar, Y. Akulova, K. M. Brown, R. Jones, D. Zhu, T. Liljeberg, and L. Liao, J. Light. Technol. **39**, 1155 (2021).
3. A. Hänsel and M. J. R. Heck, JPhys Photonics **2**, 12002 (2020).
4. X. Jia, J. Roels, R. Baets, and G. Roelkens, Sensors **21**, 5347 (2021).
5. P. T. Lin, H. G. Lin, Z. Han, T. Jin, R. Millender, L. C. Kimerling, and A. Agarwal, Adv. Opt. Mater. **4**, 1755 (2016).
6. I. E. Gordon, L. S. Rothman, C. Hill, R. V. Kochanov, Y. Tan, P. F. Bernath, M. Birk, V. Boudon, A. Campargue, K. V. Chance, B. J. Drouin, J. M. Flaud, R. R. Gamache, J. T. Hodges, D. Jacquemart, V. I. Perevalov, A. Perrin, K. P. Shine, M. A. H. Smith, J. Tennyson, G. C. Toon, H. Tran, V. G. Tyuterev, A. Barbe, A. G. Császár, V. M. Devi, T. Furtenbacher, J. J. Harrison, J. M. Hartmann, A. Jolly, T. J. Johnson, T. Karman, I. Kleiner, A. A. Kyuberis, J. Loos, O. M. Lyulin, S. T. Massie, S. N. Mikhailenko, N. Moazzen-Ahmadi, H. S. P. Müller, O. V. Naumenko, A. V. Nikitin, O. L. Polyansky, M. Rey, M. Rotger, S. W. Sharpe, K. Sung, E. Starikova, S. A. Tashkun, J. Vander Auwera, G. Wagner, J. Wilzewski, P. Wcisło, S. Yu, and E. J. Zak, J. Quant. Spectrosc. Radiat. Transf. **203**, 3 (2017).
7. R. Wang, A. Malik, I. Šimonytė, A. Vizbaras, K. Vizbaras, and G. Roelkens, Opt. Express **24**, 28977 (2016).
8. J. X. B. Sia, W. Wang, Z. Qiao, X. Li, T. X. Guo, J. Zhou, C. G. Littlejohns, C. Liu, G. T. Reed, and H. Wang, IEEE J. Quantum Electron. **56**, (2020).
9. R. Wang, S. Sprengel, A. Vasiliev, G. Boehm, J. Van Campenhout, G. Lepage, P. Verheyen, R. Baets, M.-C. Amann, and G. Roelkens, Photonics Res. **6**, 858 (2018).
10. S. A. Miller, M. Yu, X. Ji, A. G. Griffith, J. Cardenas, A. L. Gaeta, and M. Lipson, Optica **4**, 707 (2017).
11. P. Munoz, P. W. L. Van Dijk, D. Geuzebroek, M. Geiselmann, C. Dominguez, A. Stassen, J. D. Domenech, M. Zervas, A. Leinse, C. G. H. Roeloffzen, B. Gargallo, R. Banos, J. Fernandez, G. M. Cabanes, L. A. Bru, and D. Pastor, IEEE J. Sel. Top. Quantum Electron. **25**, (2019).
12. D. Z. Garbuzov, R. U. Martinelli, H. Lee, R. J. Menna, P. K. York, L. A. DiMarco, M. G. Harvey, R. J. Matarese, S. Y. Narayan, and J. C. Connolly, Appl. Phys. Lett. **70**, 2931 (1998).
13. D. Z. Garbuzov, H. Lee, V. Khalfin, R. Martinelli, J. C. Connolly, and G. L. Belenky, IEEE Photonics Technol. Lett. **11**, 794 (1999).
14. C. Lin, M. Grau, O. Dier, and M. C. Amann, Appl. Phys. Lett. **84**, 5088 (2004).
15. M. Geiselmann, https://doi.org/10.1117/12.2577588 **11689**, 116890D (2021).
16. "Photonics Simulation Software | Ansys Lumerical," https://www.ansys.com/products/photonics.
17. K. Oda, N. Takato, and H. Toba, J. Light. Technol. **9**, 728 (1991).
18. Dominik Gerhard Rabus, (Springer Berlin Heidelberg, 2007).
19. B. Liu, A. Shakouri, and J. E. Bowers, Appl. Phys. Lett. **79**, 3561 (2001).